\documentclass{iau}
\usepackage{graphicx}
\usepackage{sidecap}

\title[Microarcsecond structure via Interstellar Scintillation] 
{Microarcsecond structure of an AGN Jet via Interstellar Scintillation}

\author[H.~Bignall et al.]   
{
Hayley Bignall$^1$,
Jean-Pierre Macquart$^{1,2}$,
Leith Godfrey$^3$\thanks{Presenter.}, \\
Jeffrey Hodgson$^4$
 \and
David Jauncey$^{5,6}$
}

\affiliation{
$^1$ICRAR/Curtin University, Australia \\ 
email: {\tt H.Bignall@curtin.edu.au} \\
[\affilskip]
$^2$ARC Centre of Excellence for All-Sky Astrophysics (CAASTRO), Australia; $^3$ASTRON, Dwingeloo, The Netherlands; $^4$Max Planck Institute f\"ur Radioastronomie, Bonn, Germany; $^5$CSIRO Astronomy \& Space Science, Australia; $^6$Research School of Astronomy \& Astrophysics, ANU, Australia \\
}

\pubyear{2014}
\volume{313}
\pagerange{xx}
\jname{Extragalactic jets from every angle}
\editors{F. Massaro, C.C. Cheung, E. Lopez, A. Siemiginowska, eds.}

\begin{document}

\maketitle

\begin{abstract}
We have used the broadband backend available at the ATCA to study the fast interstellar scintillation (ISS) of quasar PKS 1257$-$326, resolving the core shift as a function of frequency on scales less than 10 microarcseconds. 
In this short paper we discuss the jet direction implied from the microarcsecond-scale core shift in PKS 1257$-$326. 
\end{abstract}

\vspace{-0.1in}

\firstsection 
\section{The ``core shift'' effect in AGN, and core shifts via ISS}~\label{sec-ISSshift}
The core of an AGN radio jet is usually the brightest, most compact feature. It is associated with the $\tau_\nu = 1$ surface, where optical depth approaches unity (\cite[Blandford \& K\"onigl 1979]{Blandford79}). The position of the core is frequency dependent due to positional variation of opacity in the jet and/or surrounding medium. Measuring the frequency dependence of the core shift provides information on the structure and physical conditions in sub-parsec scale jets, as well as the confinement mechanism and pressure gradients in the external medium. Frequency dependent core shifts can be measured with VLBI down to scales of $\sim 0.1$ mas, or with interstellar scintillation (ISS) down to micro-arcsecond scales.

For a scattering screen at distance $L$ from the observer, a frequency-dependent core shift $\Delta\theta$ will displace the scintillation patterns observed at different frequencies by a distance $L\Delta\theta$, in the opposite direction to the shift on the sky (Figure 1; see e.g.\ \cite[Little \& Hewish 1966]{Little66}). The scale of ISS in weak scattering is determined by the angular size of the first Fresnel zone, $\theta_F$, typically of order 10\,$\mu$as for scattering screens in the range $\sim 10-100$\,pc which produce rapid ISS of AGN at frequencies of several GHz.  
It is possible to measure core shifts to a precision of a fraction of the scintillation scale.

\begin{figure}
\vspace{-5mm}
 \centering
 \includegraphics[width=0.56\textwidth]{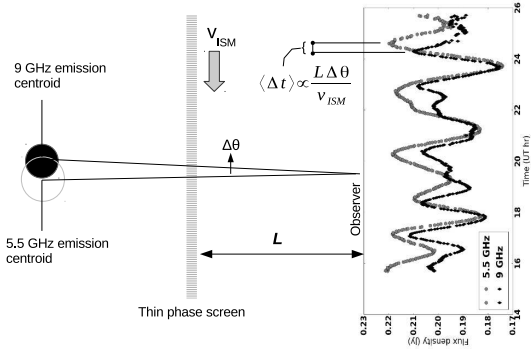} 
\vspace{-0.1in}
 \caption{Illustration of the core shift effect on the ISS pattern. Two actual light curves from PKS~1257$-$326 observed on 15 January 2011 are shown. The patterns are not identical due to increased source size relative to $\theta_F$ towards lower frequencies, but a systematic displacement $\langle \Delta t \rangle$ is measurable by cross-correlation. $\Delta t$ also depends on anisotropy in the ISS pattern. Over the full frequency range (4.5-6.5 and 8-10 GHz) we find a smooth dependence of the core-shift on frequency (\cite[Macquart et al., 2013]{Macquart13}).}
\end{figure}

\vspace{-0.1in}

\section{Frequency dependent structure of PKS~1257$-$326 via ISS}
PKS~1257$-$326 shows rapid fluctuations at centimetre wavelengths, with typical peak-to-trough variations of $\sim 20$\% in total flux density on timescales of less than 1 hour. These fluctuations are known to be ISS due to a scattering screen at a distance of $\sim 10$\,pc, based on a repeated annual cycle in characteristic timescale and pattern arrival time delays of several minutes measured between the VLA and the ATCA (\cite[Bignall et al. 2006]{Bignall06}). 
\cite{Macquart13} derived a core shift frequency dependence for PKS~1257$-$326 of $r_{core} \propto \nu^{-1/k_r}$ where $k_r > 3$, much shallower than the $\nu^{-1}$ dependence typically found in VLBI studies (e.g. \cite[Sokolovsky et al. 2011]{Sokolovsky11}). The combined fit to frequency dependence of core shift and source size, the latter determined from the timescale of ISS, implies that the jet opening angle increases with distance from the core. The results are consistent with a hydrostatically confined jet traversing a steep pressure gradient. This may help to explain the bright, relatively stable ultracompact jet in PKS 1257$-$326, implied by the presence of rapid ISS persisting now over more than a decade.

From earlier simultaneous observations at the VLA and the ATCA (\cite[Bignall et al. 2006]{Bignall06}), the ISS pattern is seen first at the VLA and then several minutes later at the ATCA, with the magnitude of the time delay depending on the time of year. Thus, the scintillation velocity is such that the ISS pattern traverses a telescope on Earth from an easterly direction; the exact direction varying over the course of a year. 
Variations at higher frequencies always lead those at lower frequencies, as observed in all data taken since 2001 (\cite[Bignall \& Hodgson 2012]{Bignall12}). Therefore, the lower frequency ISS pattern is displaced to the east of the higher frequency pattern, implying that lower frequency source component centroids are displaced to the {\em west} of the core (\S\ref{sec-ISSshift}). Taking best-fit solutions for the scintillation parameters previously determined by \cite{Bignall06} from the annual cycle in characteristic timescale and the two-station time delays, the magnitude of the core shift between 4.5 and 10 GHz is $\sim 20 \mu$as, $<0.2$\,pc at the source redshift of $z=1.256$ (\cite[Macquart et al., 2013]{Macquart13}). Although constraining the precise direction of the $\mu$as-scale core shift using further data is work in progress, the alignment is roughly consistent with the milliarcsecond- and arcsecond-scale jet components resolved directly with interferometry, which extend to the northwest of the core (\cite[Ojha et al. 2010]{Ojha10}, \cite[Bignall et al. 2006]{Bignall06}). 

\vspace{0.05in}


The Australia Telescope Compact Array is part of the Australia Telescope National Facility which is funded by the Commonwealth of Australia for operation as a National Facility managed by CSIRO.

\end{document}